\documentclass[journal]{IEEEtran}
\IEEEoverridecommandlockouts

\usepackage{color}
\usepackage{graphicx}
\usepackage{amsmath}
\usepackage{amssymb}
\usepackage{algorithm}
\usepackage{algorithmic}
\usepackage{amsmath}
\usepackage{multirow}
\usepackage{booktabs}
\usepackage{array}
\usepackage{amsthm}
\usepackage{stfloats}
\usepackage{caption}
\usepackage{subfigure}
\usepackage{bm}
\usepackage{epstopdf}
\usepackage{setspace}
\usepackage{gensymb}
\usepackage{url}
\usepackage{verbatim}

\usepackage{hyperref}
\usepackage{cleveref} 
\hypersetup{colorlinks = true, 
    linkcolor = black, 
    urlcolor = blue,
    citecolor = black} 

\newcommand{\be}{\begin{equation}}
        \newcommand{\ee}{\end{equation}}
\newcommand{\bea}{\begin{eqnarray}}
        \newcommand{\eea}{\end{eqnarray}}
\newcommand{\ba}{\begin{array}}
        \newcommand{\ea}{\end{array}}

\allowdisplaybreaks[4]

\title{Low Range-Doppler Sidelobe ISAC Waveform Design: A Low-Complexity Approach
\thanks{P. Li and M. Li are with the School of Information and Communication Engineering, Dalian University of Technology, Dalian 116024, China (e-mail: lipeishi@mail.dlut.edu.cn; mli@dlut.edu.cn).}
\thanks{R. Liu and A. Lee Swindlehurst are with the Center for Pervasive Communications and Computing, University of California, Irvine, CA 92697, USA (e-mail: rangl2@uci.edu; swindle@uci.edu).}
\thanks{Q. Liu is with the School of Computer Science and Technology, Dalian University of Technology, Dalian 116024, China (e-mail: qianliu@dlut.edu.cn).}
}
\author{Peishi Li,~\IEEEmembership{Graduate Student Member,~IEEE,}
    Ming Li,~\IEEEmembership{Senior Member,~IEEE,}
    Rang Liu,~\IEEEmembership{Member,~IEEE,} \\
    Qian Liu,~\IEEEmembership{Member,~IEEE,}
    and A. Lee Swindlehurst, ~\IEEEmembership{Fellow,~IEEE}}

\begin{document}

\maketitle
\thispagestyle{empty}
\begin{abstract}
    Integrated sensing and communication (ISAC) is a pivotal enabler for next-generation wireless networks. A key challenge in ISAC systems lies in designing dual-functional waveforms that can achieve satisfactory radar sensing accuracy by effectively suppressing range-Doppler sidelobes. However, existing solutions are often computationally intensive, limiting their practicality in multi-input multi-output (MIMO) orthogonal frequency division multiplexing (OFDM) ISAC deployments. This paper presents a novel low-complexity algorithm leveraging the augmented Lagrangian method (ALM) and Riemannian conjugate gradient (RCG) optimization techniques to address these challenges. The proposed algorithm achieves superior sidelobe suppression compared to state-of-the-art methods while dramatically reducing computational complexity, making it highly suitable for real-world MIMO-OFDM ISAC systems. Simulation results demonstrate that the proposed approach not only outperforms existing benchmarks in sidelobe reduction but also accelerates convergence, ensuring efficient performance across communication and sensing tasks.
\end{abstract}


\section{Introduction}
The rapid advancement of intelligent network applications has driven an increasing demand for wireless systems capable of supporting both high-speed communication and precise environmental sensing. This rising demand positions integrated sensing and communication (ISAC) as a pivotal technology for future sixth-generation (6G) wireless networks \cite{LiuF_JSAC_2022}, \cite{XuJindan_InfSci_2023}. By utilizing shared resources for communication and sensing, ISAC enhances spectrum efficiency, reduces hardware costs, and improves overall system performance \cite{AndrewZhang_JSTSP_2021}, \cite{LiuRang WCM 2023}.

Recent research on ISAC has predominantly focused on waveform design to balance the trade-offs between communication and sensing performance. Traditional ISAC waveform designs prioritize spatial beamforming, optimizing metrics such as signal-to-interference-plus-noise ratio (SINR) \cite{Chen_JSAC_2022}, \cite{LiuRang JSTSP 2022}, radar beampattern matching errors \cite{Tang_SAM_2020}, \cite{LiuR_JSTSP_2021}, and the Cram{\'e}r-Rao bound \cite{Liu TSP 2022}, \cite{LiuRang TWC 2024}. While these approaches have provided notable improvements in radar sensing performance, they often neglect the impact of the randomness introduced by the transmitted information, particularly on the range-Doppler sidelobes of the ambiguity function. Strong range-Doppler sidelobes will severely impair radar performance by increasing false alarms and masking weak targets \cite{Musa_arxiv_2024}, \cite{Wang_TGRS_2023}.

To address this challenge, our recent work \cite{Li_TWC_2025} introduced a symbol-level precoding (SLP)-based waveform design approach for multi-input multi-output (MIMO) orthogonal frequency division multiplexing (OFDM) ISAC systems. This approach leverages the temporal degrees of freedom (DoFs) available from SLP to effectively suppress range-Doppler sidelobes. Specifically, the study employed the majorization-minimization (MM) framework in conjunction with the alternating direction method of multipliers (ADMM) to solve the complicated waveform design problem. The objective is to minimize the range-Doppler integrated sidelobe level (ISL) while ensuring constraints on target directional power, multi-user communication quality of service (QoS), and constant-modulus transmission.

Although the MM-ADMM algorithm proposed in \cite{Li_TWC_2025} demonstrates effective sidelobe suppression, its high computational complexity limits its practicality for realistic MIMO-OFDM systems. The iterative nature of the MM-ADMM algorithm, along with the corresponding large-scale subproblems that must be solved, leads to slow convergence and significant computational overhead. Moreover, when the value of the objective function becomes small, the surrogate function constructed by the MM method may inadequately approximate the original objective, further impeding convergence and reducing accuracy. These limitations motivate the need for a more computationally efficient algorithm to realize effective sidelobe suppression.

To achieve this goal, we propose a novel low-complexity waveform design algorithm for sidelobe suppression in MIMO-OFDM systems, leveraging the augmented Lagrangian method (ALM) and the Riemannian conjugate gradient (RCG) technique. The proposed approach offers substantial improvements in both computational efficiency and sidelobe suppression performance. In contrast to MM-ADMM, the ALM-RCG algorithm integrates the constraints directly into the objective function using the augmented Lagrangian framework, reformulating the problem as an unconstrained optimization. By leveraging Riemannian manifold optimization, the algorithm efficiently explores the solution space while ensuring the constant-modulus constraints are preserved, thereby improving both the accuracy and computational efficiency of the optimization process. Simulation results demonstrate that the proposed ALM-RCG approach not only significantly reduces computational complexity compared to the existing MM-ADMM method but also achieves superior sidelobe suppression, making it well-suited for practical deployment in MIMO-OFDM ISAC systems.

\section{System Model and Problem Formulation}
\subsection{Transmit Signal Model}
In this paper, we consider a MIMO-OFDM ISAC system, where a dual-function base station (BS) equipped with $ N_{\text{t}} $ transmit antennas emits an OFDM signal comprising $ N_{\text{c}} $ subcarriers and $ N_{\text{s}} $ symbols to serve $ K $ single-antenna communication users while simultaneously performing radar sensing\footnote[1]{To focus on the design of waveforms with low ambiguity function sidelobes, we assume that the self-interference due to full-duplex operation is effectively suppressed by means of advanced self-interference cancellation techniques \cite{He_JSAC_2023}.}. Let $\mathbf{s}_{n, m} \triangleq \left[ s_{n, m, 1}, \dots, s_{n, m, K} \right]^T \in \mathbb{C}^{K} $ denote the modulated symbol vector on the $n$-th subcarrier during the $ m $-th symbol slot.
To effectively leverage multi-user interference and suppress the range-Doppler sidelobes of the ambiguity function, the dual-function BS employs a nonlinear SLP technique that accounts for both the modulated symbols within each time slot and their temporal characteristics across slots.
The corresponding symbol-level precoded signal is denoted as $ \mathbf{x}_{n, m} \in \mathbb{C}^{N_{\text{t}}} $. The baseband OFDM transmit signal is then expressed as
\begin{equation} \label{eq:baseband_OFDM}
    \widetilde{\mathbf{x}}(t) = \frac{1}{\sqrt{N_{\text{c}}}}  \sum_{m = 0}^{N_{\text{s}}-1} \sum_{n = 0}^{N_{\text{c}}-1}  \mathbf{x}_{n, m} e^{\jmath 2 \pi n \Delta f t_m} \text{rect} \bigg( \frac{t_m}{T_{\text{tot}}} \bigg),
\end{equation}
where we define the following variables: $\Delta f = 1/T$ is the subcarrier spacing, $T$ is the OFDM symbol duration, $T_{\text{tot}} = T + T_{\text{CP}}$ is the total symbol duration, $T_{\text{CP}}$ is the portion of $T_{\text{tot}}$ due to the cyclic prefix (CP), and $t_m=t- m T_{\text{tot}}$ is the relative fast-time variable during the $m$-th OFDM symbol. The function $\text{rect}(t)$ is a rectangular pulse that takes the value 1 for $t \in [0, 1]$ and $0$ otherwise.

\subsection{Multi-user Communication Model}
Assuming that the communication channels experience frequency-selective fading, the received signal on the $n$-th subcarrier for the $k$-th user is given by
\begin{equation} \label{eq:received signal}
    y_{n, m, k} = \mathbf{h}_{n, k}^H \mathbf{x}_{n,m} + z_{n, m, k},
\end{equation}
where $\mathbf{h}_{n, k} \in \mathbb{C}^{N_{\text{t}}}$ denotes the corresponding frequency-domain channel that is assumed to be known at the BS, and $z_{n, m, k} \sim \mathcal{CN} (0, \sigma^2)$ is additive white Gaussian noise (AWGN). Unlike conventional block-level precoding schemes, where the precoded signal $\mathbf{x}_{n,m}$ is obtained through a linear transformation of the symbol vector $\mathbf{s}_{n,m}$, SLP leverages a nonlinear mapping to directly optimize  $\mathbf{x}_{n,m}$ based on both the instantaneous channel state information (CSI) and the specific symbols $\mathbf{s}_{n,m}$ to be transmitted to all the users. By exploiting the symbol-wise structure of the modulation constellation, SLP actively shapes the transmitted signal at the symbol level, enabling enhanced interference exploitation and constructive interference (CI) maximization.

To simplify the description of SLP, we assume that the modulated symbol $s_{n, m, k}$ is generated using $\Omega$-phase-shift-keying (PSK) ($\Omega$ = 2, 4, \dots). The safety margin associated with the transmit symbol $s_{n, m, k}$ is defined as the distance between the noiseless received signal $\mathbf{h}_{n, k}^H \mathbf{x}_{n, m}$ and the nearest decision boundary, expressed as \cite{LiuR_JSAC_2022}
\begin{equation}\label{eq:communication constraint}
    \delta_{n, m, k} = \mathfrak{R}\{ \widetilde{y}_{n, m, k} \} \sin \phi - \big| \mathfrak{I}\{ \widetilde{y}_{n, m, k} \} \big| \cos \phi,
\end{equation}
where $\widetilde{y}_{n, m, k} = \mathbf{h}_{n, k}^{H} \mathbf{x}_{n, m} s_{n, m, k}^{\ast}$ denotes the rotated noiseless received signal, and $\phi = \pi / \Omega$. A larger safety margin moves the received signal further away from the decision boundaries, reducing the likelihood of noise-induced errors. Consequently, the multi-user communication QoS constraints are formulated as
\begin{equation}
    \mathfrak{R}\{ \widetilde{y}_{n,m,k} \}  \sin \phi   -\big|\mathfrak{I}\{ \widetilde{y}_{n,m,k} \}\big| \cos \phi \geq \gamma,
\end{equation}
where $\gamma$ is the desired minimum safety margin. To facilitate the algorithm development, we apply fundamental linear algebra principles to reformulate (\ref{eq:communication constraint}) into an equivalent compact form:
\begin{equation}\label{eq:comm_constraint}
    \mathfrak{R}\big\{ \widetilde{\mathbf{h}}^H_{j} \mathbf{x} \big\} \geq \gamma, ~~ \forall j = 1, 2, \dots, 2 K N_{\text{s}} N_{\text{c}},
\end{equation}
where we define
\begin{subequations}
    \begin{align}
        \mathbf{x}_{m}                                            & \triangleq \big[ \, \mathbf{x}_{0,m}^T, \mathbf{x}_{1,m}^T, \dots, \mathbf{x}^T_{N_{\text{c}}-1,m} \, \big]^T, \\
        \mathbf{x}                                                & \triangleq \big[ \, \mathbf{x}^T_0, \mathbf{x}^T_1, \dots, \mathbf{x}^T_{N_{\text{s}}-1} \, \big]^T ,          \\
        \!\! \widetilde{\mathbf{h}}_{2K(n+mN_{\text{c}})+ 2k}^H   & \triangleq \mathbf{e}_{i}^T \otimes \mathbf{h}_{n, k}^H s_{n,m,k}^{\ast} ( \sin\phi -\jmath \cos\phi ),        \\
        \!\! \widetilde{\mathbf{h}}_{2K(n+mN_{\text{c}})+ 2k-1}^H & \triangleq \mathbf{e}_{i}^T \otimes \mathbf{h}_{n, k}^H s_{n,m,k}^{\ast} (\sin\phi + \jmath \cos\phi ),
    \end{align}
\end{subequations}
$i = 1+n+mN_{\text{c}}$, and $\mathbf{e}_{i} \in \mathbb{R}^{N_{\text{s}}N_{\text{c}}}$ indicates the $i$-th column of an $N_{\text{s}}N_{\text{c}} \times N_{\text{s}}N_{\text{c}}$ identity matrix. In the remainder of the paper,  (\ref{eq:comm_constraint}) will be used as the multi-user communication performance constraint.

\subsection{Radar Sensing Model}
In radar signal analysis and waveform design, the range-Doppler sidelobe level of the ambiguity function serves as a critical metric for assessing target detection and parameter estimation performance.
The ambiguity function represents the time-frequency composite auto-correlation of the transmitted signal and is defined as
\begin{equation} \label{eq:AF_define}
    \chi(\tau,f_{\text{d}}) \triangleq \int_{-\infty}^{\infty} \widetilde{x}_0(t) \widetilde{x}_0^{\ast}(t+\tau)e^{\jmath 2 \pi f_{\text{d}}t} \text{d}t,
\end{equation}
where $\tau$ is the round-trip time delay, $f_{\text{d}}$ is the Doppler shift, $\widetilde{x}_0(t)=\mathbf{a}_{\text{T}}^H(\theta_0) \widetilde{\mathbf{x}}(t)$ is the OFDM signal after beamforming to the known azimuth angle $\theta_0$. The vector $\mathbf{a}_{\text{T}}(\theta) \in \mathbb{C}^{N_{\text{t}}}$ is the transmit steering vector, which is given by
\begin{equation}
    \mathbf{a}_{\text{T}}(\theta_0) \triangleq \big[ 1, e^{\jmath 2 \pi  \sin(\theta_0) \frac{d_{\text{T}}}{\lambda} }, \dots, e^{\jmath 2 \pi (N_{\text{t}}-1) \sin(\theta_0) \frac{d_{\text{T}}}{\lambda} }  \big]^T,
\end{equation}
where $d_{\text{T}}$ denotes the transmit antenna spacing, and $\lambda$ is the wavelength.

Given the presence of a CP in the OFDM waveform, we can use the discrete periodic ambiguity function of the OFDM signal to simplify the derivation \cite{Li_TWC_2025}, \cite{Mercier_TAES_2020}, which can be written as
\setcounter{equation}{10}
\begin{subequations}\label{eq:AF}
    \begin{align}
        \chi(l, \nu)
         & = \sum_{m = 0}^{N_{\text{s}}-1} \sum_{n = 0}^{N_{\text{c}}-1} \mathbf{x}_{n,m}^H \mathbf{A}\mathbf{x}_{n,m} e^{-\jmath 2 \pi l\frac{n}{N_{\text{c}}}} e^{\jmath 2\pi \nu \frac{m}{N_{\text{s}}}} \\
         & =\! \sum_{m = 0}^{N_{\text{s}}-1} \mathbf{x}_m^H \widetilde{\mathbf{A}}   \mathbf{D}^{\ast}_{l} \widetilde{\mathbf{A}}^H \mathbf{x}_m  e^{\jmath 2\pi \nu \frac{m}{N_{\text{s}}}}                \\
         & = \mathbf{x}^H \widetilde{\mathbf{A}} \big( \mathbf{D}_{\nu} \otimes \mathbf{D}^{\ast}_{l} \big) \widetilde{\mathbf{A}}^H \mathbf{x}, \label{eq:AF_waveform}
    \end{align}
\end{subequations}
where we define
\begin{subequations}\label{eq:def_Matrix}
    \begin{align}
        \mathbf{A}       & \triangleq \mathbf{a}_{\text{T}}(\theta_0) \mathbf{a}_{\text{T}}^H(\theta_0) , ~~~~~ \widetilde{\mathbf{A}} \triangleq \mathbf{I}_{N_{\text{s}}N_{\text{c}}} \otimes \mathbf{a}_{\text{T}}(\theta_0), \label{eq:A_tilde_def} \\
        \mathbf{D}_{l}   & \triangleq \text{diag}\big( 1, e^{-\jmath2 \pi l \frac{1}{N_{\text{c}}}}, \dots, e^{-\jmath2 \pi l \frac{N_{\text{c}}-1}{N_{\text{c}}}} \big),  \label{eq:D_l_define}                                                        \\
        \mathbf{D}_{\nu} & \triangleq \text{diag}(1, e^{\jmath2 \pi \nu \frac{1}{N_{\text{s}}}}, \dots, e^{\jmath 2 \pi \nu  \frac{N_{\text{s}}-1}{N_{\text{s}}}}).  \label{eq:D_nu_define}
    \end{align}
\end{subequations}

Ideally, the ambiguity function should exhibit a ``thumbtack'' shape, characterized by a narrow mainlobe peak and low sidelobes. However, the shape of the ambiguity function is heavily influenced by the dual-function waveform  $\mathbf{x}$, whose inherent randomness can result in elevated range-Doppler sidelobes. Therefore, careful design of the dual-function waveform is essential to effectively suppress these sidelobes. To quantify the sidelobe levels of the ambiguity function, the ISL is commonly employed as the performance metric \cite{Lihusheng_WCNC_2022}:
\begin{equation} \label{eq:ISL}
    \xi_{\text{ISL}} \triangleq  \sum_{(l, \nu) \in \mathcal{R}_{\text{s}}} \Big| \mathbf{x}^H \widetilde{\mathbf{A}} \big( \mathbf{D}_{\nu} \otimes \mathbf{D}^{\ast}_{l} \big) \widetilde{\mathbf{A}}^H \mathbf{x} \Big|^2,
\end{equation}
where $\mathcal{R}_{\text{s}} = \big\{ (l, \nu) \big| ~ |l| \leq N-1, |\nu| \leq M-1, (l, \nu) \neq (0, 0) \big\}$ denotes the sidelobe region.

In addition to range-Doppler sidelobes, the radar receive signal-to-noise ratio (SNR) is critical for ensuring high sensing performance. In the absence of precise information about the target range and radar cross section (RCS), the target directional power at $\theta_0$ is used as an alternative metric to ensure a satisfactory radar receive SNR, expressed as \cite{Li_TWC_2025}
\begin{subequations}\label{eq:IL_power}
    \begin{align}
        P_{\text{IL}}
         & = \sum_{m = 0}^{N_{\text{s}}-1} \mathbf{x}_{m}^H (\mathbf{F}_{N_\text{c}} \otimes \mathbf{I}_{N_{\text{t}}})  (\mathbf{I}_{N_{\text{c}}} \otimes \mathbf{A} ) (\mathbf{F}_{N_\text{c}}^H \otimes \mathbf{I}_{N_{\text{t}}}) \mathbf{x}_{m} \\
         & = \sum_{m = 0}^{N_{\text{s}}-1} \mathbf{x}_{m}^H  (\mathbf{I}_{N_{\text{c}}} \otimes \mathbf{A} ) \mathbf{x}_{m}                                                                                                                           \\
         & = \mathbf{x}^H \overline{\mathbf{A}} \mathbf{x},
    \end{align}
\end{subequations}
where $\overline{\mathbf{A}} = \mathbf{I}_{N_{\text{s}}N_{\text{c}}} \otimes \mathbf{A}$, and $\mathbf{F}_{N_\text{c}} \in \mathbb{C}^{N_\text{c} \times N_\text{c}}$ denotes the normalized discrete Fourier transform (DFT) matrix.

\newcounter{TempEqCnt1}
\setcounter{TempEqCnt1}{\value{equation}}
\setcounter{equation}{8}
\begin{figure*}[!t]
    \begin{equation} \label{eq:alm}
        g(\widetilde{\mathbf{x}}) =
        \sum_{(l, \nu) \in \mathcal{R}_{\text{s}}} \big| \widetilde{\mathbf{x}}^H \mathbf{B}_{l, \nu} \widetilde{\mathbf{x}} \big|^2 + \frac{\rho}{2} \max \big\{ P_0 - \widetilde{\mathbf{x}}^H \widehat{\mathbf{A}} \widetilde{\mathbf{x}} + \mu / \rho, 0 \big\}^2
        + \frac{\rho}{2} \sum_{j=1}^{2KN_{\text{s}}N_{\text{c}}} \max \big\{  \gamma - \mathfrak{R}\big\{ \widetilde{\mathbf{h}}_{j}^H \widetilde{\mathbf{F}} \widetilde{\mathbf{x}} \big\} + \lambda_j / \rho, 0   \big\}^2.
    \end{equation}
    \begin{equation}\label{eq:gradx}
        \begin{aligned}
             & \nabla g(\widetilde{\mathbf{x}}) = 2 \sum_{(l, \nu) \in \mathcal{R}_{\text{s}}} \big(  \widetilde{\mathbf{x}}^H \mathbf{B}_{l, \nu} \widetilde{\mathbf{x}} \mathbf{B}_{l, \nu}^H \widetilde{\mathbf{x}} + \widetilde{\mathbf{x}}^H \mathbf{B}_{l, \nu}^H \widetilde{\mathbf{x}} \mathbf{B}_{l, \nu} \widetilde{\mathbf{x}}  \big) + 2 \rho
            \begin{cases}
                (\widetilde{\mathbf{x}}^H \widehat{\mathbf{A}} \widetilde{\mathbf{x}} - P_0 + \mu / \rho ) \widehat{\mathbf{A}}\widetilde{\mathbf{x}} & ~ P_0 -\widetilde{\mathbf{x}}^H \widehat{\mathbf{A}} \widetilde{\mathbf{x}} + \mu / \rho > 0     \\
                ~ \mathbf{0}                                                                                                                          & ~ P_0 - \widetilde{\mathbf{x}}^H \widehat{\mathbf{A}} \widetilde{\mathbf{x}} + \mu / \rho \leq 0
            \end{cases}
            \\
             & \hspace{5.7cm} +  \rho \sum_{j=1}^{2KN_{\text{s}}N_{\text{c}}}
            \begin{cases}
                \big( \mathfrak{R}\big\{ \widetilde{\mathbf{h}}_{j}^H \widetilde{\mathbf{F}} \widetilde{\mathbf{x}} \big\} - \gamma - \lambda_j / \rho \big) \widetilde{\mathbf{F}}^H \widetilde{\mathbf{h}}_{j} & \gamma - \mathfrak{R}\big\{ \widetilde{\mathbf{h}}_{j}^H \widetilde{\mathbf{F}} \widetilde{\mathbf{x}} \big\} + \lambda_j / \rho   > 0.  \\
                ~ \mathbf{0}                                                                                                                                                                                     & \gamma - \mathfrak{R}\big\{ \widetilde{\mathbf{h}}_{j}^H \widetilde{\mathbf{F}} \widetilde{\mathbf{x}} \big\} + \lambda_j / \rho \leq 0.
            \end{cases}
        \end{aligned}
    \end{equation}
    \rule[-0pt]{18.1 cm}{0.05em}
\end{figure*}
\setcounter{equation}{\value{TempEqCnt1}}

\subsection{Problem Formulation}
Based on the above discussion, our objective is to design the SLP transmit waveform $\mathbf{x}$ to minimize the range-Doppler ISL, subject to constraints on target directional power, multi-user communication QoS, and constant-modulus transmission. The corresponding SLP-based ISAC waveform design problem is formulated as
\begin{subequations} \label{eq:original_problem}
    \begin{align}
         & \underset{\mathbf{x}}{\min} ~~  \sum_{(l, \nu) \in \mathcal{R}_{\text{s}}} \big| \mathbf{x}^H  \widetilde{\mathbf{A}} \big( \mathbf{D}_{\nu} \otimes \mathbf{D}^{\ast}_{l} \big)\widetilde{\mathbf{A}}^H  \mathbf{x} \big|^2 \\
         & ~ \text{s.t.} \hspace{12pt}  \mathbf{x}^H  \overline{\mathbf{A}} \mathbf{x} \geq P_0  , \label{eq:pro1_IL_constraint}                                                                                                        \\
         & \hspace{27pt} \mathfrak{R}\big\{ \widetilde{\mathbf{h}}_{j}^H \mathbf{x} \big\} \ge \gamma, ~~ \forall j = 1, \dots, 2 K N_{\text{s}} N_{\text{c} } ,                                                                        \\
         & \hspace{27pt} \big| \widetilde{\mathbf{F}}^H \mathbf{x} \big| = \sqrt{P_{\text{T}} / N_{\text{tot}}} \mathbf{1}_{N_{\text{tot}}}, \label{eq:pro1_cm_constraint}
    \end{align}
\end{subequations}
where $P_0$ is the required minimum target directional power, $P_{\text{T}}$ is the transmit power budget, $N_{\text{tot}} = N_{\text{s}}N_{\text{c}}N_{\text{t}}$, $\widetilde{\mathbf{F}} = \mathbf{I}_{N_{\text{s}}} \otimes \mathbf{F}_{N_\text{c}} \otimes \mathbf{I}_{N_{\text{t}}}$, and $\widetilde{\mathbf{x}} = \widetilde{\mathbf{F}}^H \mathbf{x}$ is the transmitted time-domain data signal.

The optimization problem  in (\ref{eq:original_problem}) poses significant challenges due to its non-convex quartic objective function and the non-convex constraints in  (\ref{eq:pro1_IL_constraint}) and (\ref{eq:pro1_cm_constraint}). While the MM-ADMM algorithm proposed in \cite{Li_TWC_2025} can be employed to address these issues, its practical implementation is hindered by the large size of the subproblems and the slow convergence rate of the algorithm, resulting in a prohibitive computational burden. To overcome these limitations, we propose a more efficient solution in the Riemannian space, referred to as the ALM-RCG algorithm, which is presented below.

\section{Low-Complexity Waveform Design Algorithm}

Considering that $\widetilde{\mathbf{F}}$ is a full-rank matrix with $\widetilde{\mathbf{F}}^{-1} = \widetilde{\mathbf{F}}^H$, and noting the relationship between the frequency-domain data $\mathbf{x}$ and the time-domain data $\widetilde{\mathbf{x}}$ is $\widetilde{\mathbf{x}} = \widetilde{\mathbf{F}}^H \mathbf{x}$, the waveform design problem (\ref{eq:original_problem}) can be reformulated as
\begin{subequations} \label{eq:refrom_problem}
    \begin{align}
         & \underset{\widetilde{\mathbf{x}}}{\min} ~~~  \sum_{(l, \nu) \in \mathcal{R}_{\text{s}}} \big| \widetilde{\mathbf{x}}^H \mathbf{B}_{l,\nu} \widetilde{\mathbf{x}} \big|^2        \\
         & ~ \text{s.t.} \hspace{12pt}  \widetilde{\mathbf{x}}^H \widehat{\mathbf{A}} \widetilde{\mathbf{x}} \geq P_0  ,  \label{eq:re_pro_IL_constraint}                                  \\
         & \hspace{27pt} \mathfrak{R}\big\{ \widetilde{\mathbf{h}}_{j}^H \widetilde{\mathbf{F}} \widetilde{\mathbf{x}} \big\} \ge \gamma, ~~ \forall j , \label{eq:re_pro_comm_constraint} \\
         & \hspace{27pt} \big| \widetilde{\mathbf{x}} \big| = \sqrt{P_{\text{T}} / N_{\text{tot}}} \mathbf{1}_{N_{\text{tot}}}, \label{eq:re_pro_cm_constraint}
    \end{align}
\end{subequations}
where we define $\mathbf{B}_{l,\nu} \triangleq \widetilde{\mathbf{F}}^H \widetilde{\mathbf{A}} \big( \mathbf{D}_{\nu} \otimes \mathbf{D}^{\ast}_{l} \big)\widetilde{\mathbf{A}}^H \widetilde{\mathbf{F}}$, $\widehat{\mathbf{A}} \triangleq \widetilde{\mathbf{F}}^H  \overline{\mathbf{A}} \widetilde{\mathbf{F}}$.

Note that the constant-modulus constraint (\ref{eq:re_pro_cm_constraint}) can be geometrically interpreted as a complex circle manifold:
\begin{equation}
    \!\!\!  \mathcal{M} = \big\{ \widetilde{\mathbf{x}} \in \mathbb{C}^{N_{\text{tot}}} : |\widetilde{x}_{v}| = \sqrt{ P_{\text{T}}/N_{\text{tot}}},  \forall v = 1, \dots, N_{\text{tot}} \big\}.
\end{equation}
Thus, we can employ ALM to incorporate the constraints of target directional power (\ref{eq:re_pro_IL_constraint}) and communication QoS (\ref{eq:re_pro_comm_constraint}) into the objective function, and transform problem (\ref{eq:refrom_problem}) into an unconstrained optimization problem on the complex circle manifold, which can be expressed as
\begin{equation} \label{eq:problem_alm}
    \underset{\widetilde{\mathbf{x}} \in \mathcal{M} }{\min} ~~~ g(\widetilde{\mathbf{x}}),
\end{equation}
where $g(\widetilde{\mathbf{x}})$ is formulated in (\ref{eq:alm}) on the top of this page, $\rho$ is a penalty parameter, and $\mu$ and $\boldsymbol{\lambda} \triangleq [\lambda_1, \dots, \lambda_{2K N_{\text{s}}N_{\text{c}}}]$ are Lagrangian dual variables. In the $t$-th iteration, the dual variables are updated by
\begin{subequations} \label{eq:update_dualvar}
    \begin{align}
        \!\! \mu^{t+1} \!         & = \min \!\big\{ \! \max \{ 0, \mu^t  \! +  \rho (P_0 - (\widetilde{\mathbf{x}}^t)^H \widehat{\mathbf{A}} \widetilde{\mathbf{x}}^t) \},  \mu_{\text{max}}\big\},     \label{eq:update_mu}                                           \\
        \!\! \lambda_{j}^{t+1} \! & = \min \!\big\{ \! \max  \{ 0, \lambda_{j}^t +  \rho  ( \gamma - \mathfrak{R} \{ \widetilde{\mathbf{h}}_{j}^H \widetilde{\mathbf{F}} \widetilde{\mathbf{x}}^t \} )  \},  \lambda_{\text{max}}\big\}, \!\! \label{eq:update_lambda}
    \end{align}
\end{subequations}
where $\mu_{\text{max}}$ and $\lambda_{\text{max}}$ are the maximum limits imposed on $\mu$ and $\lambda_j$, respectively, to mitigate the impact of ill-conditioning.

Given fixed dual variables, (\ref{eq:problem_alm}) becomes an unconstrained optimization problem on the manifold, which can be solved by a first-order RCG algorithm. The algorithm follows a series of steps: obtaining the Riemannian gradient, selecting the descent direction, determining the step size in conjunction with the descent direction, and updating the next feasible solution. By iteratively repeating these steps until convergence, we ultimately obtain the final solution.

Specifically, in the $q$-th iteration of the RCG algorithm, the Riemannian gradient $\nabla_{\mathcal{M}} g(\widetilde{\mathbf{x}}_q^t)$ at point $\widetilde{\mathbf{x}}_q^t$ is derived by projecting the Euclidean gradient $\nabla g(\widetilde{\mathbf{x}}_q^t)$ onto the tangent space as
\begin{equation} \label{eq:gradx_Riemannian}
    \begin{aligned}
        \nabla_{\mathcal{M}} g(\widetilde{\mathbf{x}}_q^t)
         & = \mathcal{P}_{T_{\widetilde{\mathbf{x}}_q^t} \mathcal{M}} \big(\nabla g(\widetilde{\mathbf{x}}_q^t)\big)                                                                            \\
         & = \nabla g(\widetilde{\mathbf{x}}_q^t) - \mathfrak{R} \big\{ \nabla g(\widetilde{\mathbf{x}}_q^t) \odot (\widetilde{\mathbf{x}}_q^t)^{\ast} \big\} \odot \widetilde{\mathbf{x}}_q^t,
    \end{aligned}
\end{equation}
where $\nabla g(\widetilde{\mathbf{x}}_q^t)$ can be obtained via (\ref{eq:gradx}), $T_{\widetilde{\mathbf{x}}_q^t} \mathcal{M} $ represents the tangent space at point $\widetilde{\mathbf{x}}_q^t$ on the manifold $\mathcal{M}$, and $\mathcal{P}_{T_{\widetilde{\mathbf{x}}_q^t} \mathcal{M}} (\mathbf{z}) = \mathbf{z} - \mathfrak{R} \{ \mathbf{z} \odot (\widetilde{\mathbf{x}}_q^t)^{\ast} \} \odot \widetilde{\mathbf{x}}_q^t$ is a projection operator. Next, we calculate the conjugate descent direction $\boldsymbol{\eta}_q$.

\begin{algorithm}[!t]
    \begin{small}
        \caption{ALM-RCG Algorithm for solving problem (\ref{eq:refrom_problem})}
        \label{alg:ALM-RCG}
        \begin{algorithmic}[1]
            \REQUIRE {$\widetilde{\mathbf{h}}_j$, $\forall j$, $\gamma$, $\widehat{\mathbf{A}}$, $\mathbf{B}_{l, \nu}$, $\forall l$, $\nu$,  $P_0$, $P_{\text{T}}$, $\alpha_{\text{th}}$, $\mu_{\text{max}}$, $\lambda_{\text{max}}$.}
            \ENSURE {$\widetilde{\mathbf{x}}^{\star}$.}
            \STATE {Initialize $\widetilde{\mathbf{x}}^0$, $\mu^0$, $\boldsymbol{\lambda}^0$, $t:=0$.}
            \REPEAT
            \STATE {Initialize $q:=0$, $\widetilde{\mathbf{x}}_0^t := \widetilde{\mathbf{x}}^t$, $\boldsymbol{\eta}_0 := - \nabla_{\mathcal{M}} g(\widetilde{\mathbf{x}}^t)$.}
            \STATE {Update $\widetilde{\mathbf{x}}_1^t = \sqrt{P_{\text{T}}/N_{\text{tot}}}e^{\jmath \angle (\widetilde{\mathbf{x}}_{0}^t + \alpha_0 \boldsymbol{\eta}_0)}$, $q:=q+1$.}
            \REPEAT
            \STATE {Calculate  the descent direction $\boldsymbol{\eta}_q$ by (\ref{eq:descent_direction}).}
            \STATE {Calculate the step size $\alpha_q$ by Armijo line search strategy.}
            \STATE {Update $\widetilde{\mathbf{x}}_{q+1}^{t}$ by (\ref{eq:updata_x}).}
            \STATE {$q:=q+1$.}
            \UNTIL {$ \alpha_q \leq \alpha_{\text{th}} $}
            \STATE {$\widetilde{\mathbf{x}}^{t+1}:= \widetilde{\mathbf{x}}_{q}^{t}$.}
            \STATE {Update Lagrangian dual variables $\mu^{t+1}$ and $\lambda_j^{t+1}$ by (\ref{eq:update_dualvar}).}
            \STATE {$t:=t+1$.}
            \UNTIL { $\| \widetilde{\mathbf{x}}^t- \widetilde{\mathbf{x}}^{t-1} \| \leq \delta_{\text{th}}$ }
            \STATE {Return $\widetilde{\mathbf{x}}^{\star}:=\widetilde{\mathbf{x}}^t$.}
        \end{algorithmic}
    \end{small}
\end{algorithm}

Although the steepest-descent direction can be used to obtain a locally optimal solution, it is known to lead to sluggish convergence in practical applications. Consequently, we employ the Polak-Ribi\'{e}re (PR) conjugate gradient method, which incorporates more gradient information and typically achieves faster convergence. In the $q$-th iteration, the descent direction of the PR conjugate gradient method is given by \cite{Absil_book_2009}
\begin{equation} \label{eq:descent_direction}
    \boldsymbol{\eta}_q = - \nabla_{\mathcal{M}} g(\widetilde{\mathbf{x}}_q^t) + \beta_q^{\text{PR}}
    \mathcal{P}_{T_{\widetilde{\mathbf{x}}_q^t} \mathcal{M}} (\boldsymbol{\eta}_{q-1}),
\end{equation}
where $\beta_q^{\text{PR}}$ is the PR update coefficient:
\begin{equation} \label{eq:pr_coefficient}
    \beta_q^{\text{PR}} \triangleq \frac{ \big \langle \nabla_{\mathcal{M}} g(\widetilde{\mathbf{x}}_q^t), \nabla_{\mathcal{M}} g(\widetilde{\mathbf{x}}_q^t) - \mathcal{P}_{T_{\widetilde{\mathbf{x}}_q^t} \mathcal{M}} (g(\widetilde{\mathbf{x}}_{q-1}^t)) \big \rangle }{ \big \langle \nabla_{\mathcal{M}} g(\widetilde{\mathbf{x}}_{q-1}^t), \nabla_{\mathcal{M}} g(\widetilde{\mathbf{x}}_{q-1}^t)  \big \rangle},
\end{equation}
and $\langle \cdot, \cdot \rangle$ denotes the matrix inner product. The step size $\alpha_q$ is obtained by the Armijo line search strategy. Finally, to ensure that the updated point $\widetilde{\mathbf{x}}_{q+1}^t$ remains on the manifold, the following retraction mapping of $\widetilde{\mathbf{x}}$ is employed:
\begin{equation} \label{eq:updata_x}
    \begin{aligned}
        \widetilde{\mathbf{x}}_{q+1}^t
         & = \mathcal{R}_{\widetilde{\mathbf{x}}_q^t} (\alpha_q \boldsymbol{\eta}_q)  = \sqrt{ P_{\text{T}}/N_{\text{tot}}} e^{\jmath \angle (\widetilde{\mathbf{x}}_q^t + \alpha_q \boldsymbol{\eta}_q )},
    \end{aligned}
\end{equation}
where $\mathcal{R}(\cdot)$ denotes a retraction operation on the manifold $\mathcal{M}$.

Based on the above derivations, the proposed ALM-RCG algorithm is summarized in Algorithm \ref{alg:ALM-RCG}, where $\alpha_{\text{th}}$ is the minimum step size allowed. Next, we analyze the computational complexity of Algorithm \ref{alg:ALM-RCG}. The dominant computational cost of the proposed ALM-RCG algorithm arises from the calculation of the Euclidean gradient (\ref{eq:gradx}), which has a complexity of $\mathcal{O} \big\{ N_{\text{tot}} (N_{\text{s}} N_{\text{c}}^2 + KN_{\text{c}} + N_{\text{t}}) \big\}$. Accordingly, the overall computational complexity of Algorithm \ref{alg:ALM-RCG} is $\mathcal{O} \big\{I N_{\text{tot}} (N_{\text{s}} N_{\text{c}}^2 + KN_{\text{c}} + N_{\text{t}}) \big\}$, where $I$ denotes the total number of required iterations. This represents a substantial reduction compared to the MM-ADMM algorithm, which has a complexity of $\mathcal{O} \big\{ I \sqrt{(2K + N_{\text{t}}) N_{\text{s}} N_{\text{c}}} N_{\text{tot}} ( N_{\text{tot}}^2 + K N_{\text{s}} N_{\text{c}} ) \big\}$ \cite{Li_TWC_2025}. Additionally, we also analyze the memory/storage requirements of the proposed ALM-RCG algorithm, which is $\mathcal{O} \big\{ N_{\text{tot}} (K N_{\text{s}} + N_{\text{t}}) + N_{\text{s}}^2 N_{\text{c}}^2 \big\}$. This is larger than that required by the MM-ADMM algorithm, which is $\mathcal{O} \big\{ N_{\text{tot}} (K + N_{\text{t}}) + N_{\text{s}}^2 N_{\text{c}}^2 \big\} $, due to the additional variables involved in the ALM-RCG algorithm.

\section{Simulation Results}
In the following simulations, we consider a MIMO-OFDM ISAC BS equipped with $ N_{\text{t}} = 6 $ antennas arranged as a uniform linear array with half-wavelength spacing. The BS serves $ K = 4 $ users using an OFDM waveform with $ N_{\text{c}} = 32 $ subcarriers and blocks of $ N_{\text{s}} = 16 $ OFDM symbols. The noise power is set to $ \sigma^2 = -70 $dBm, the transmit power is $ P_{\text{T}} = 10 $W, and the required minimum target directional power is $ P_0 = 8$W. The desired minimum safety margin is defined as $\gamma \triangleq \sigma \sin \phi \sqrt{\Gamma}$, where $\Gamma=10$ dB is the SINR threshold. Additionally, the convergence threshold of the ALM-RCG algorithm is set as $\delta_{\text{th}} = 10^{-4}$. The radar receiver is assumed to perform a two-dimensional DFT to obtain the range-Doppler map, and target detection and parameter estimation are performed by searching for peaks in the range-Doppler map \cite{Musa_arxiv_2024}, \cite{Li_TWC_2025}.

\begin{figure}[!t]
    \centering
    \vspace{-1mm}
    \hspace{-2mm}
    \includegraphics[width = \linewidth]{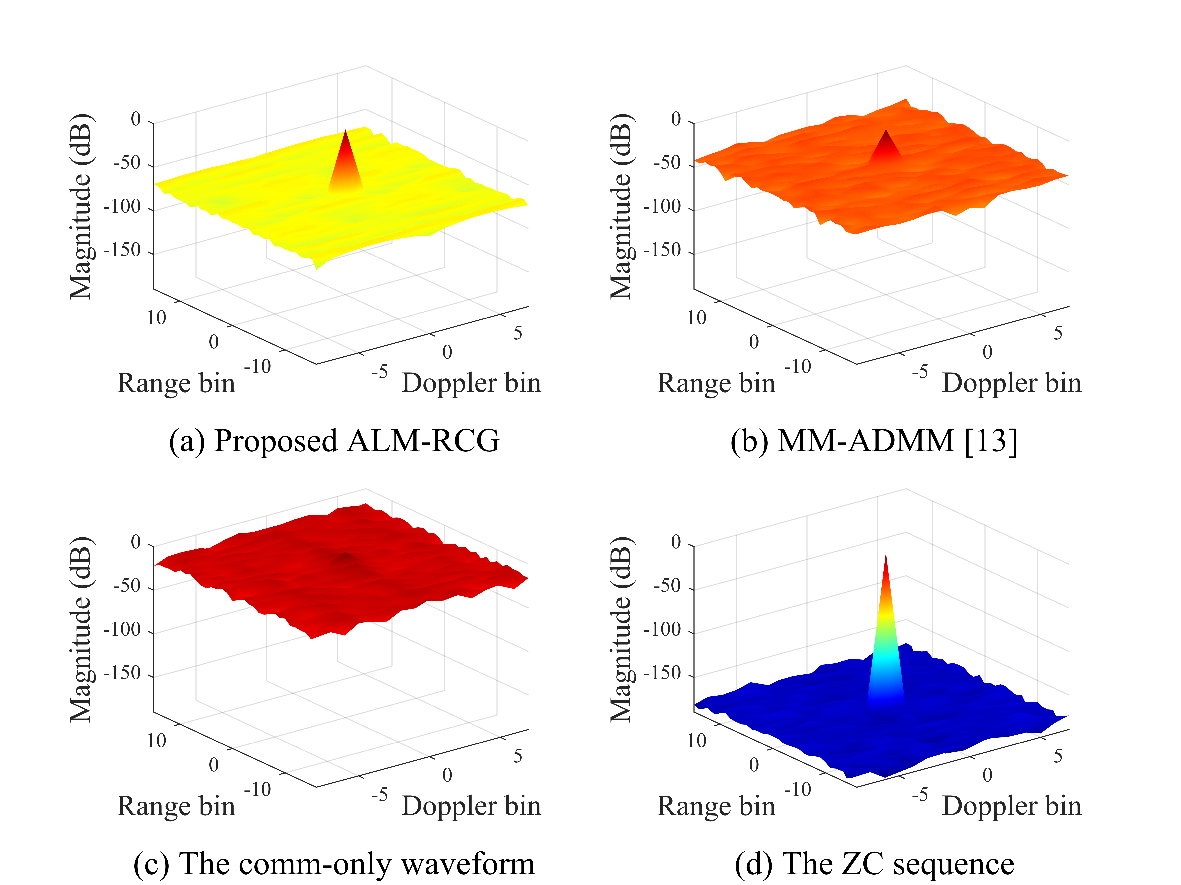}
    \caption{The ambiguity functions of different waveforms.}
    \label{fig:ambiguity_function}
\end{figure}

\begin{figure}[!t]
    \centering
    \vspace{-1mm}
    \includegraphics[width = \linewidth]{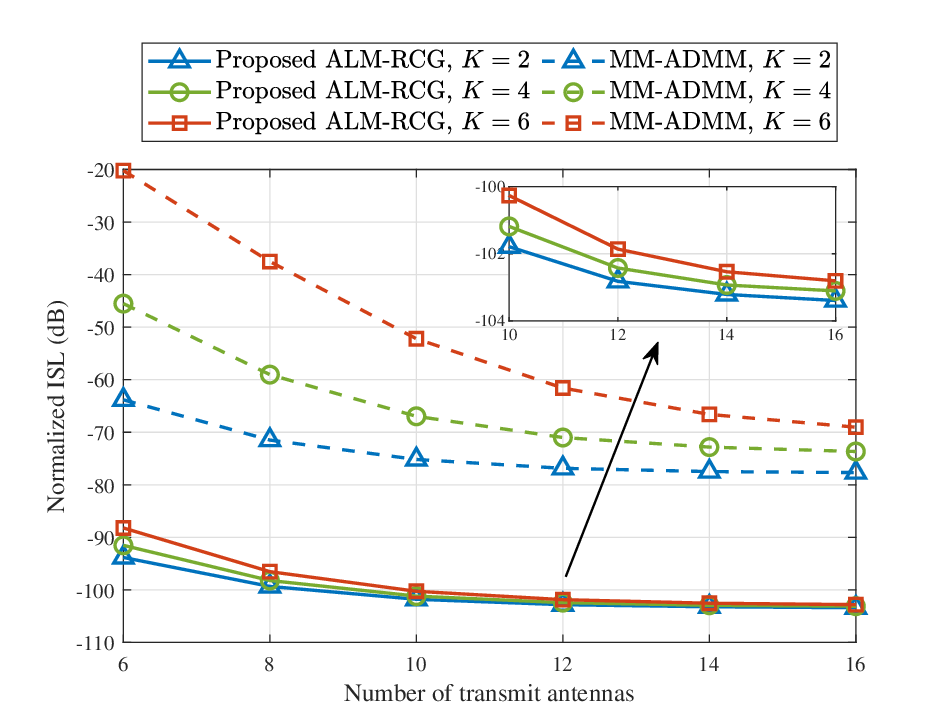}
    \caption{The normalized range-Doppler ISL versus the number of transmit antennas.}
    \label{fig:ISLR_Nt}\vspace{-3mm}
\end{figure}

We begin by evaluating the range-Doppler sidelobe suppression performance through the ambiguity functions of different waveform designs, as shown in Fig. \ref{fig:ambiguity_function}. To assess the effectiveness of the proposed ALM-RCG algorithm, we include comparisons with the SLP-based ISAC waveform obtained using the ``\textbf{MM-ADMM}'' algorithm \cite{Li_TWC_2025}, the SLP-based ``\textbf{Comm-only}'' waveform obtained by solving the conventional communication max-min fairness problem, and a radar-only waveform using the ``\textbf{Zadoff-Chu (ZC) sequence}'' \cite{Chu_TIT_1972}. The ZC sequence with length $N$ is defined as $r_{p}[n] = e^{-\jmath \pi p \frac{n(n+1)}{N}}$, where $0 \leq n < N$, and the root index $p$ is an integer relatively prime to $N$. As illustrated in Fig. \ref{fig:ambiguity_function}, the proposed ALM-RCG algorithm achieves significantly lower range-Doppler sidelobes compared to not only the comm-only waveform, but also the previously proposed MM-ADMM approach. This improvement is attributed to the fact that, for small values of the range-Doppler ISL objective function, the surrogate function employed by the MM algorithm may poorly approximate the original function, negatively impacting both convergence speed and accuracy.

\begin{figure}[!t]
    \centering
    \vspace{-0mm}
    \includegraphics[width = \linewidth ]{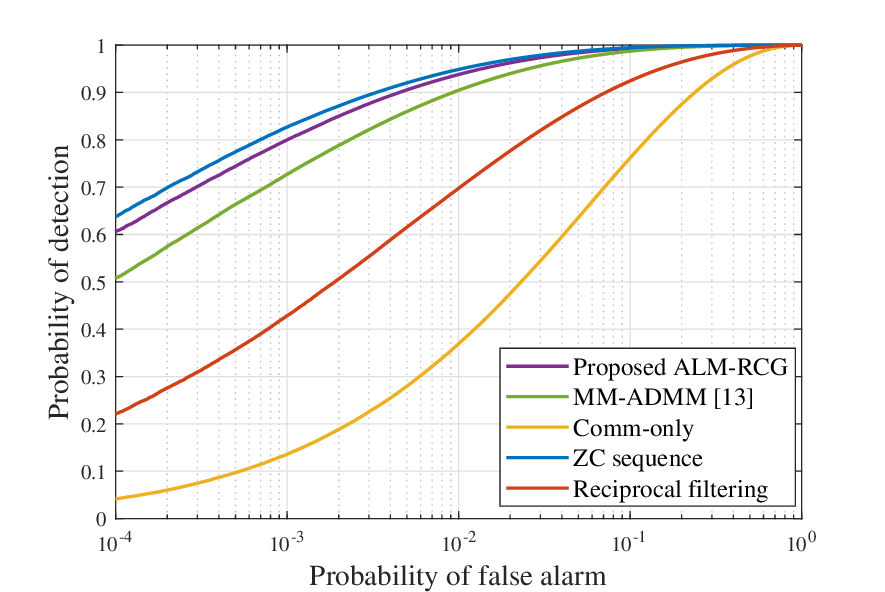}
    \vspace{-1mm}
    \caption{ROC curve for various algorithms.}
    \label{fig:ROC}
    \vspace{-3mm}
\end{figure}

\begin{figure}[!t]
    \centering
    \subfigbottomskip=-4pt
    \subfigcapskip=-2pt
    \vspace{-2mm}
    \subfigure[Range estimation RMSE.]{
        \includegraphics[width=  \linewidth]{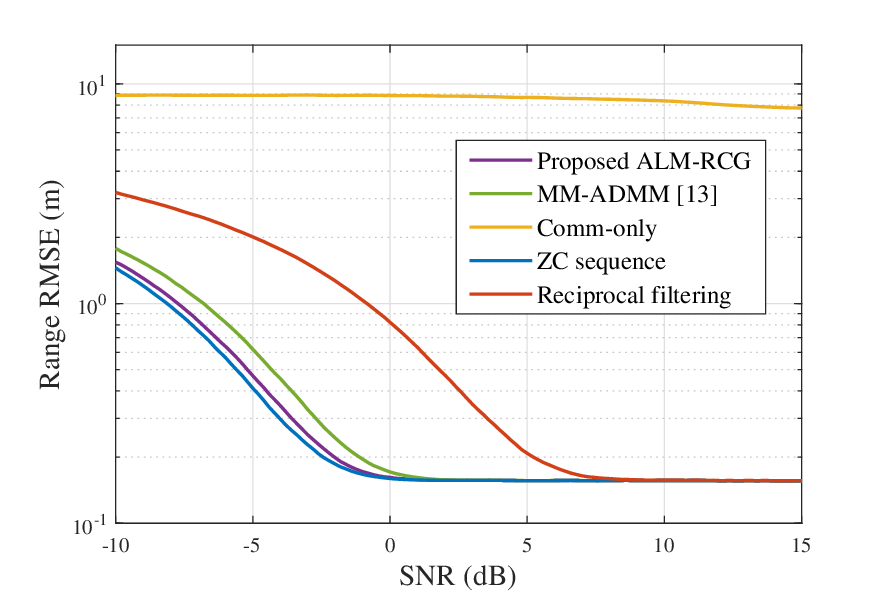} \label{fig:range_RMSE}
    }
    \subfigure[Velocity estimation RMSE.]{
        \includegraphics[width= \linewidth]{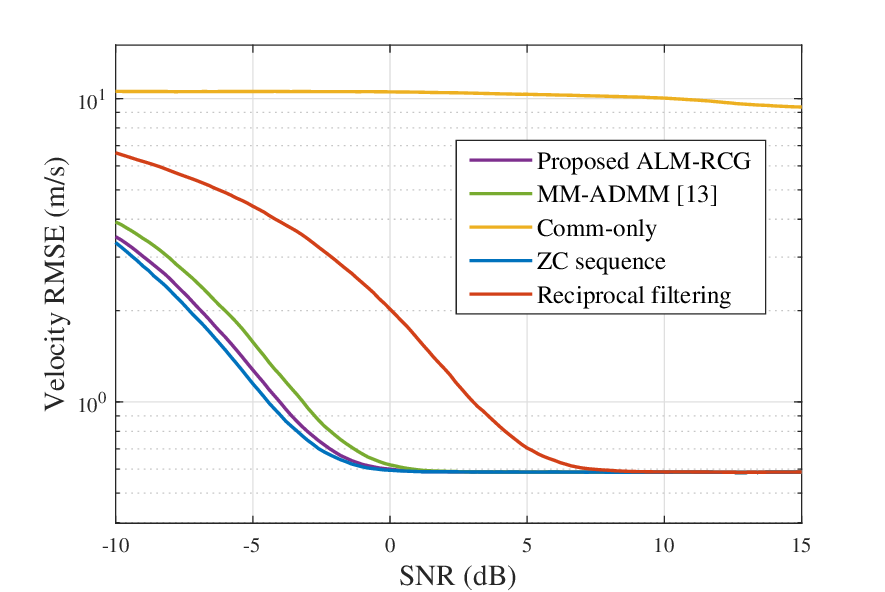} \label{fig:velo_RMSE}
    }
    \centering
    \caption{RMSE for range and velocity estimates of the weak target versus the sensing SNR.}\label{fig:RMSE}
    \vspace{-3mm}
\end{figure}

Fig. \ref{fig:ISLR_Nt} illustrates the normalized range-Doppler ISL versus the number of transmit antennas for different numbers of users. We see that the proposed ALM-RCG algorithm significantly reduces the range-Doppler ISL of the ambiguity function, outperforming the MM-ADMM algorithm. As the number of antennas increases, the additional DoFs available for waveform design enable further sidelobe suppression. Note also that the range-Doppler ISL increases with the number of communication users, highlighting the inherent trade-off between sensing and communication performance.

Next, we present the receiver operating characteristic (ROC) curves for the various algorithms in Fig. \ref{fig:ROC}, including the ``reciprocal filtering'' method from \cite{Sturm_IEEEProc_2011} for comparison. The ROC of ALM-RCG closely approaches the performance upper bound achieved by the radar-only ZC sequence and significantly outperforms the other methods. Fig. \ref{fig:RMSE} illustrates the root-mean-square error (RMSE) of the range and velocity estimates for various waveforms as a function of the sensing SNR. To achieve the same RMSE performance, the proposed ALM-RCG algorithm requires $1$ dB lower SNR than MM-ADMM and approximately $7$ dB less than the reciprocal filtering method. Furthermore, the proposed ALM-RCG algorithm achieves nearly identical range and velocity RMSE performance as that of the radar-only ZC sequence.

\begin{figure}[!t]
    \centering
    \includegraphics[width = \linewidth]{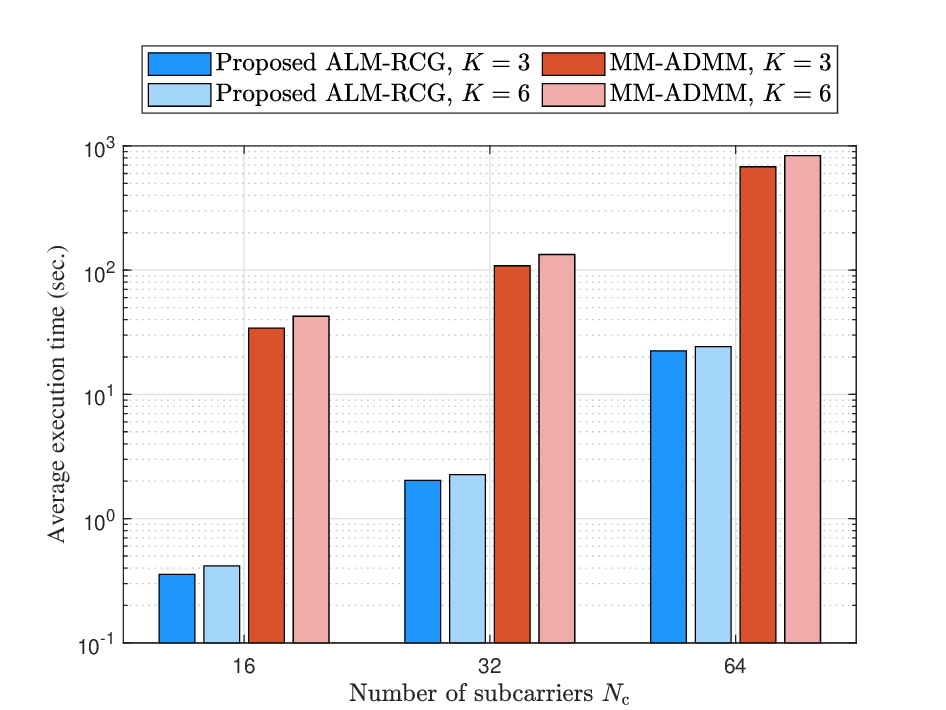}
    \caption{The average execution time under different numbers of users and subcarriers.}
    \label{fig:execu_time}
    \vspace{-3mm}
\end{figure}

Finally, in Fig. \ref{fig:execu_time} we evaluate the computational complexity of the ALM-RCG and MM-ADMM algorithms based on the average execution time required to generate the transmit waveform. The results clearly demonstrate that ALM-RCG significantly outperforms MM-ADMM in terms of computational efficiency across all tested configurations. In particular, ALM-RCG reduces the execution time by a factor of approximately 30 to 100, depending on the number of users and subcarriers. This substantial improvement in computational efficiency comes at the cost of increased storage requirements, with ALM-RCG requiring approximately 1.33 to 2.65 times the storage of MM-ADMM under the experimental settings shown in Fig. \ref{fig:execu_time}. Moreover, the execution time of the proposed ALM-RCG algorithm exhibits a moderate increase with the number of users and a more pronounced increase with the number of subcarriers, which is consistent with the theoretical computational complexity analysis of Section III. These findings highlight the superior efficiency and scalability of ALM-RCG, demonstrating its potential as a computationally efficient and robust solution for ISAC waveform design.

\vspace{-0 mm}

\section{Conclusions}
This paper has investigated the SLP-based ISAC waveform design problem for range-Doppler sidelobe suppression in MIMO-OFDM systems, and proposed a low-complexity solution that leverages the ALM and RCG techniques. The proposed approach effectively overcomes the computational challenges associated with the existing state-of-the-art SLP-based algorithm, while still achieving sidelobe suppression and target parameter estimation performance similar to waveform designs that consider only the radar functionality.

\end{document}